\documentclass[11pt]{article}

\usepackage[margin=1in]{geometry}
\usepackage{lmodern}
\usepackage[T1]{fontenc}
\usepackage[utf8]{inputenc}
\usepackage{amsmath}
\usepackage{amssymb}
\usepackage{graphicx}
\usepackage{booktabs}
\usepackage{array}
\usepackage{seqsplit}
\usepackage{xurl}
\usepackage{microtype}
\usepackage{xcolor}
\usepackage{listings}
\usepackage[numbers,sort&compress]{natbib}
\usepackage[hidelinks]{hyperref}

\hypersetup{
  colorlinks=true,
  linkcolor=black,
  urlcolor=blue!60!black,
  citecolor=blue!60!black,
  pdftitle={A measurement substrate for agentic Kubernetes operations},
  pdfauthor={Joshua Odmark, Gideon Rubin, Deon van der Vyver},
  pdfsubject={Methodology paper: closed-loop measurement of autonomous Kubernetes operators},
  pdfkeywords={agentic systems, software engineering methodology, fault injection, Kubernetes, retrieval-augmented agents, falsification}
}

\makeatletter
\let\origtexttt\texttt
\renewcommand{\texttt}[1]{\origtexttt{\bgroup\hyphenchar\font=`\-\relax #1\egroup}}
\makeatother
\newcommand{\codebreak}[1]{\texttt{\seqsplit{#1}}}

\title{
  A measurement substrate for agentic Kubernetes operations\\[0.3em]
  \large Methodology and a case study in retrieval-compounding falsification
}

\author{
  Joshua Odmark\thanks{Independent. Correspondence: \href{mailto:joshua.odmark@gmail.com}{joshua.odmark@gmail.com}.}
  \and Gideon Rubin\thanks{LDE.}
  \and Deon van der Vyver\thanks{Cognyx.}
}

\date{May 2026}

\begin{document}

\maketitle
\thispagestyle{empty}

\sloppy

\begin{abstract}
Empirical claims about autonomous Kubernetes operations agents are largely unfalsifiable. Published work reports observational results without controlled comparisons against an agent-disabled baseline, selection bias is endemic, pre-registered decision matrices are absent, and samples are typically too small for the noise level of the underlying scoring system. The cause is the same gap that limits the agents themselves: code agents have a verification substrate that turns ``did it work'' into a fast, falsifiable, ground-truth signal, and operations has nothing equivalent. We present agent-breakage, a closed-loop measurement framework that injects faults into a target Kubernetes cluster, observes how an autonomous agent responds, scores the response on four axes against ground truth, and accumulates outcome-labeled (state, action, outcome) tuples. The framework distinguishes framework error from reasoning error, supports a true off-condition control via a deterministic-embedder mechanism, and enforces pre-registered decision matrices. We use it as a case study to test whether retrieval over past postmortems compounds an agent's capability. The methodological payload is three confounds the substrate caught during that case study, each of which would have produced a wrong published claim on a less instrumented version of the same work: a pgvector index bug, a $+19\%$ selection-bias artifact, and small-sample estimates that overstated effects by roughly $3\times$. The retrieval result itself is a partial falsification: 1 of 3 dense-corpus scenarios significant at $p<0.05$, pooled effect $+3.9$ percentage points, not significant at $n=60$. A within-scenario corpus-density sweep at 360 runs shows that mechanistic alignment of near-neighbors dominates raw count. The framework is released open source.
\end{abstract}

\section{Introduction}

Empirical claims about autonomous operations agents are, in their current state, largely unfalsifiable. The deployed-agent literature (K8sGPT, HolmesGPT, Robusta, and adjacent projects) is rich in capability demonstrations and sparse in controlled measurement. Where numbers appear, they are typically observational, drawn from production runs against production failures the agent has already seen. Selection bias is endemic. Pre-registered controlled comparisons against an agent-disabled baseline are rare. Sample sizes routinely sit below what the underlying scoring system's noise level can support.

The gap is methodological as much as instrumental, and the contrast with code agents is illustrative. Code agents work because the test suite is the agent's external memory of correctness. A failing test is unambiguous; a passing one is overwhelming evidence. The agent doesn't reason from priors about whether its change worked; it observes the answer. ``Did it work, with cluster-derived ground truth'' is the affordance under most published advances in agentic software engineering.

Operations has nothing equivalent. The closest analogs in cluster operations (canary rollouts, shadow traffic, blue-green deploys) are approximate and reactive. They tell an operator whether something has already gone wrong; they do not give an agent the \emph{fast, parallel, speculative} property that turns ``will this fix work'' into ``I observed it work in eight forks before it touched production.'' Without that substrate, autonomous operations agents reason about cluster behavior from priors, and empirical claims about how those agents behave reason from incident-history-shaped samples drawn from production. Most observable production failures of agent autonomy in operations contexts trace to the first gap; most of the field's unfalsifiable claims trace to the second.

This paper presents a closed-loop measurement framework, \texttt{agent-breakage}, designed to attach falsifiable numbers to claims about Kubernetes operations agents. The framework deliberately injects faults into a target Kubernetes cluster, observes how an autonomous agent responds, scores the response on four axes against ground truth defined per scenario, and accumulates structured (state, action, outcome) tuples that feed retrieval-augmented inference on the agent's next incident. It distinguishes \emph{framework error} (substrate bugs) from \emph{reasoning error} (agent mistakes), so infrastructure failures do not corrupt agent-capability claims and vice versa.

We use the framework to test one empirical hypothesis: \emph{retrieval over past postmortems compounds the agent's capability over time}. The hypothesis is the central argument for why a substrate of this shape is worth building. If past experience doesn't measurably improve future agent behavior, the substrate is gathering data with no consumer.

The decision matrix (continue / ship-limited / pivot) was written down before any run. The control arm filters all retrieved content out via a deterministic-embedder mechanism, which makes ``retrieval off'' actually retrieval off rather than retrieval with noise. Sample sizes match the noise level of a $[0,1]$-bounded four-axis score.

The headline result is a partial falsification. At single-cluster corpus density, retrieval delivers a statistically significant positive effect ($\Delta=+0.058$, $p<0.05$) on the densest-corpus scenario tested, weaker positive trends on a second, and a slightly negative trend on a third. Pooled across three dense-corpus scenarios at $n=60$, the effect is $+0.039$ percentage points, not significant (\href{https://github.com/odmarkj/agent-breakage/blob/v0.1.0/breakage/reports/falsification-test-2026-04-24.md}{falsification-test-2026-04-24.md}). A within-scenario corpus-density sweep (360 runs, $n=60$ per density tier) confirms that \emph{per-scenario heterogeneity dominates per-density variance}: mechanistic alignment of near-neighbors matters more than count (\href{https://github.com/odmarkj/agent-breakage/blob/v0.1.0/breakage/reports/corpus-density-sweep-2026-04-28.md}{corpus-density-sweep-2026-04-28.md}). An $n=40$ rerun of a near-significant scenario from the original test confirms that the original $t=1.82$ was noise; the underlying effect is small or zero (\href{https://github.com/odmarkj/agent-breakage/blob/v0.1.0/breakage/reports/n40-rerun-2026-04-28.md}{n40-rerun-2026-04-28.md}).

The strong version of the compounding hypothesis (\emph{more postmortems, more retrieval at inference time, monotonically better agent}) does not survive at the corpus density a single cluster can produce.

The retrieval finding isn't the result that transfers most. The substrate's behavior during the work that produced it is. Three confounds got caught and corrected during the experimental sequence, each of which would have produced a wrong published result on a less instrumented version of the same work: a \texttt{pgvector} ivfflat index bug that returned sporadic empty result sets, a $+19\%$ selection-bias artifact mistaken for a causal retrieval effect, and small-sample estimates that overstated effect sizes by roughly $3\times$. The substrate's \emph{capacity for self-correction} is what distinguishes a measurement framework from a demonstration tool. It's the methodology contribution worth centering on.

All numbers reported here are reproducible from a clean machine in approximately five hours of wall clock per arm; the framework, the agent under test, the scenarios, and the reproducer harnesses are released under Apache 2.0 at the \texttt{v0.1.0} tag.

\subsection{Contributions}

This paper makes three contributions:

\begin{enumerate}
\item \textbf{Three methodological pathologies caught and corrected during agentic-operations measurement work} (\S4). Any agentic-operations measurement framework built on observational corpora and small samples will face the same three traps. This is the contribution that generalizes most cleanly across the broader agentic-evaluation literature, and the right primary criterion for evaluating agentic-operations measurement frameworks.

\item \textbf{A closed-loop measurement framework for Kubernetes operations agents} that surfaced those pathologies (\S3). The framework, the agent under test (Emily), the scenario library, and the experimental harnesses are publicly released at \href{https://github.com/odmarkj/agent-breakage}{github.com/odmarkj/agent-breakage} under Apache 2.0. The reproducibility tag \texttt{v0.1.0} resolves to the commit producing the reported numbers.

\item \textbf{A pre-registered, controlled, partial falsification of the retrieval-compounding hypothesis at single-cluster corpus density}, presented as the case study that exercised the framework (\S5). The result is statistically significant on one of three dense-corpus scenarios; mixed on the others. A within-scenario corpus-density sweep refines the binding constraint from ``density'' to ``mechanistic alignment,'' a finding that has implications for retrieval-augmented agents in causal domains more broadly (\S6).
\end{enumerate}

\subsection{Roadmap}

\S2 surveys the (sparse) related work in agentic-operations measurement and explains why fixed-ground-truth benchmarks cannot catch the confounds this work catches. \S3 describes the substrate's architecture, scoring rules, and the framework-error/reasoning-error distinction. \S4 documents the three substrate-caught methodological pathologies and argues they are the durable contribution. \S5 presents the retrieval-compounding case study, including the controlled falsification experiment, the corpus-density sweep, and the $n=40$ reruns. \S6 discusses what the corpus-density sweep reveals about semantic vs mechanistic retrieval. \S7 lists limitations and explicit scope of what this paper is and is not claiming. \S8 concludes. Appendices give reproducibility instructions, the controlled vocabulary, and the full scenario inventory.

\section{Related work}

This work sits at the intersection of four threads: LLM-augmented operations tooling, agentic systems evaluation, retrieval-augmented agents, and cluster-fault injection. We survey each briefly and identify what is missing across them: a measurement substrate that produces falsifiable claims about agent behavior under controlled fault conditions on running infrastructure.

\subsection{LLM-augmented operations tooling}

A small ecosystem of open-source tools applies large language models to Kubernetes operations. K8sGPT~\citep{k8sgpt} (CNCF Sandbox; Alex Jones et al.) provides LLM-driven diagnostic explanations for cluster issues, scanning resources and surfacing human-readable analyses. HolmesGPT~\citep{holmesgpt} (Robusta) extends the pattern toward alert investigation and root-cause triage, using an LLM to follow runbooks and propose diagnoses. Robusta itself~\citep{robusta} predates the LLM-augmented variant as an alert-management and incident-response platform; the LLM integration is a recent layer.

Each of these tools is useful as deployed; each ships without a measurement layer rigorous enough to falsify its own claims. Reported ``accuracy'' or ``helpfulness'' numbers are typically observational, drawn from in-production runs, and lack a controlled comparison against an LLM-disabled baseline. None publishes pre-registered decision matrices, deterministic-control arms, or sample sizes appropriate for the underlying score's noise level. This is not a critique of the tools, they are genuinely useful, but it is the gap our framework is designed to address. Without a substrate that distinguishes \emph{the agent helped} from \emph{the incident was easy}, comparative claims about agentic-operations capability remain unfalsifiable.

\subsection{Agentic systems evaluation}

Recent benchmarks evaluate large language models acting as agents across a range of environments. AgentBench~\citep{agentbench} (Liu et al., 2023) covers eight environments including operating-system command execution, database queries, and web browsing. SWE-bench~\citep{swebench} (Jimenez et al., 2023) tests an agent's ability to resolve real GitHub issues by producing patches that pass the issue's tests; SWE-bench Verified is OpenAI's curated subset. OSWorld~\citep{osworld} (Xie et al., 2024) benchmarks multimodal agents in real desktop computer environments. ToolLLM~\citep{toolllm} (Qin et al., 2023) evaluates agents on real-world API usage at scale.

Each of these is rigorous within its scope, and each has shaped our understanding of agent capability. None evaluates agents against running infrastructure with deliberately injected faults, ground-truth fault states, and accumulating outcome-labeled tuples. The closest comparison, the OS environment in AgentBench, uses scripted shell-task evaluation, not live cluster-state measurement with reversibility-aware tooling and cluster-derived ground truth. The substrate we present is closer in shape to the SWE-bench framing (real-world tasks, reproducible test harness) than to AgentBench's scripted environments, but applied to the operations domain that none of the existing benchmarks cover.

\subsection{Retrieval-augmented agents}

Retrieval-augmented generation~\citep{rag} (Lewis et al., 2020) and its successors have shaped how language models incorporate external context. The agent-specific variants are particularly relevant. ReAct~\citep{react} (Yao et al., 2022) interleaves reasoning traces with tool actions, with retrieval as one possible action. Reflexion~\citep{reflexion} (Shinn et al., 2023) gives agents verbal feedback from prior failures, accumulating a kind of episodic memory. Voyager~\citep{voyager} (Wang et al., 2023) builds a skill library from successful Minecraft trajectories, exhibiting open-ended capability growth as the library compounds.

The compounding mechanism we test, retrieval over past postmortems improving future agent behavior, is analogous to Voyager's skill-library compounding, but applied to a domain (operations) where ground truth is cluster-state-derived rather than goal-state-derived. The retrieval-augmentation literature is largely textual: corpora are documents, queries return passages, and the agent reasons over the passages' text. Our case is unusual in that the retrieved items are \emph{outcome-labeled action tuples}, past postmortems with \texttt{resolved} / \texttt{regressed} / \texttt{inconclusive} labels and the agent's actual action sequences. We have not found a published retrieval-augmented-agent benchmark that uses outcome-labeled action tuples as the corpus on infrastructure-level operations tasks. The closest comparison is the skill-library line of work, where the corpus is similarly action-shaped but the domain is open-ended (Minecraft) rather than constrained to a controlled-vocabulary fault space.

\subsection{Cluster-fault injection}

Chaos engineering for Kubernetes is a mature space. Chaos Mesh~\citep{chaosmesh} (CNCF Incubating; PingCAP) provides a CRD-driven framework for injecting pod, network, IO, and time faults. LitmusChaos~\citep{litmus} (CNCF Incubating) offers a similar surface organized around chaos experiments and chaos hubs. Gremlin~\citep{gremlin} is the commercial counterpart, with an opinionated control plane and SaaS deployment.

These tools are designed for resilience testing, verifying that \emph{systems} withstand specific failure modes. They are not paired with agent measurement; the chaos-engineering layer and the agent-evaluation layer are historically separate concerns. Our framework reuses the same conceptual primitive (deliberate fault injection) but layers it under a single closed-loop measurement system that scores an \emph{agent's response} to the fault rather than the system's intrinsic resilience. The injectors in our framework are simpler than Chaos Mesh's (the scope is one fault per scenario, not failure-mode composition), and could be replaced by Chaos Mesh CRDs in a future integration. The novelty is not the injection, which is well-explored; the novelty is the closed loop from injection through agent action through outcome-labeled tuple back to retrieval at the next incident.

\subsection{Why fixed-ground-truth benchmarks cannot catch these confounds}

The benchmarks surveyed in \S2.2 are rigorous within scope, but each operates on a ground truth that is fixed at evaluation time. SWE-bench grades a candidate patch against a pre-existing test suite; the test suite either passes or fails, and the candidate task pool is locked at corpus-construction time. AgentBench and ToolLLM resolve agent behavior into scripted pass/fail decisions over fixed task pools. OSWorld grades against deterministic screenshot-equality targets. Across all four, the population of tasks the agent is evaluated on is the same population the next agent in the leaderboard is evaluated on, and the per-task verdict is a deterministic function of agent output and a fixed scorer.

Fixed ground truth makes the selection bias discussed in \S4.2 of this paper structurally impossible. If an agent retrieves an ``easier'' subset of the task pool, the next baseline is evaluated on the same pool, and the comparison cancels. There is no incident-history-shaped population for retrieval to select from differentially.

Operations differs in kind, not degree. Ground truth in this paper's framework moves \emph{during} measurement: the workload shifts, the agent learns from its own postmortems, the incident population drifts as cluster state evolves. Past failures do not predict future failure distributions because each new scenario run mutates the corpus that the next scenario will retrieve against. A code-agent benchmark with this property would resemble a SWE-bench whose test suite was being rewritten between submissions by the agents being evaluated.

This is why none of the four benchmarks above could surface the three pathologies the substrate caught. Their construction guarantees the kind of fixed-population determinism that makes those pathologies invisible. The substrate's contribution is to make agentic-operations measurement work under moving ground truth: detect that motion, and expose the confounding effects it would otherwise hide. Code-agent benchmarking solves a different problem with a different shape; the techniques in this paper are not redundant with theirs.

\section{The measurement substrate}

The framework is a closed loop. Each iteration runs a \emph{scenario}, a YAML file describing one specific cluster fault, the conditions for considering it fixed, the conditions for considering an out-of-scope regression, and the ground-truth root cause from a controlled vocabulary. A scenario run executes the full loop end-to-end:

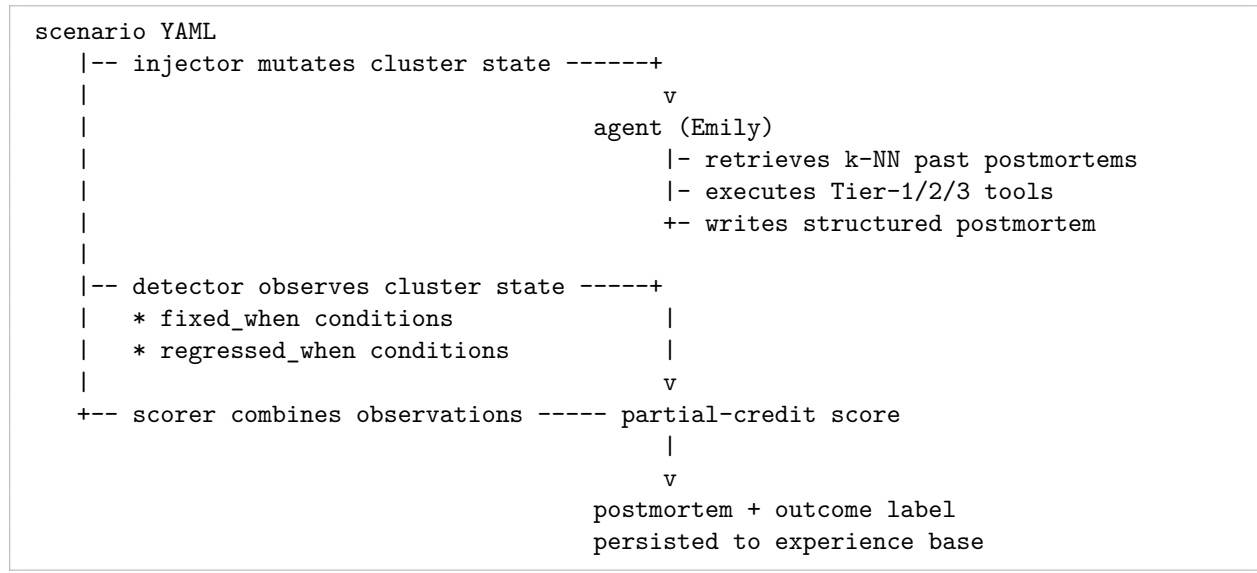
\begin{lstlisting}
 scenario YAML
    |-- injector mutates cluster state ------+
    |                                         v
    |                                    agent (Emily)
    |                                         |- retrieves k-NN past postmortems
    |                                         |- executes Tier-1/2/3 tools
    |                                         +- writes structured postmortem
    |
    |-- detector observes cluster state -----+
    |   * fixed_when conditions               |
    |   * regressed_when conditions           |
    |                                         v
    +-- scorer combines observations ----- partial-credit score
                                              |
                                              v
                                         postmortem + outcome label
                                         persisted to experience base
\end{lstlisting}

The agent's tool calls are wrapped by a \emph{speculative-execution controller} that snapshots cluster state before any Tier-2 mutation, watches for SLO regression in the seconds that follow, and auto-reverts on regression. The controller produces a mechanical revert reason that the agent reads on its next inference cycle; it does not itself reason semantically about why the metric moved (\texttt{breakage/src/speculative-exec/}; \href{https://github.com/odmarkj/agent-breakage/blob/v0.1.0/operator/docs/speculative-execution.md}{operator/docs/speculative-execution.md}).

Retrieval over past postmortems happens \emph{before} the agent's first tool call. Top-k similar postmortems, with their outcome labels (\texttt{resolved}, \texttt{regressed}, \texttt{inconclusive}), are injected into the agent's context as exemplars. Both resolved and regressed exemplars are surfaced; filtering out failures would lose counterexample signal. Retrieval is the only mechanism through which past experience influences current behavior.

\subsection{Components}

The substrate decomposes into seven components, each with a narrow contract:

\textbf{Runner.} A Fastify HTTP server orchestrating scenario execution. Exposes endpoints for run dispatch (\texttt{POST /run}), retrieval (\texttt{POST /retrieve}), postmortem capture (\texttt{POST /capture-postmortem}), and mid-investigation hypothesis capture (\texttt{POST /capture-hypothesis}). Single-active-scenario model; multi-tenant execution would require additional registry work. (\texttt{breakage/src/runner/})

\textbf{Injectors.} One module per fault type, each implementing a thin \texttt{Injector} interface that returns an \texttt{Undo} thunk run at scenario end. Implementations include \texttt{deployment-patch}, \texttt{secret-content}, \texttt{configmap-patch}, \texttt{flagd-flag} (for OpenTelemetry-Demo workloads), \texttt{network-policy}, and \texttt{pod-evict}. Adding a new injector is the most common framework extension; each is $\leq 200$ lines. (\texttt{breakage/src/injector/})

\textbf{Detectors.} Evaluate \texttt{fixed\_when} and \texttt{regressed\_when} expressions per scenario, dispatching across handlers for the Kubernetes API and Prometheus. Conditions can opt into \texttt{skip\_if\_unevaluable: true} to treat unreachable backends (e.g., Prometheus down in a k3d setup) as pass. Each condition has an optional \texttt{sustained\_for\_s} window; the condition must hold continuously for that window to pass, eliminating transient-blip false positives. (\texttt{breakage/src/detector/})

\textbf{Scorer.} A pure-logic module that combines the observation, the agent's postmortem, the agent's mid-investigation hypotheses, and the retrieved exemplars into a single \texttt{ScoreResult}. The scoring rules are detailed in \S3.2. (\texttt{breakage/src/scorer/})

\textbf{Experience base.} PostgreSQL with \texttt{pgvector}. One table, \texttt{postmortems}, indexed via HNSW (\texttt{m=16, ef\_construction=64}). The default embedder is \texttt{OpenAICompatibleEmbedder} against an in-cluster \texttt{text-embeddings-inference} (TEI) pod serving \texttt{BAAI/bge-m3} (1024-dim). A swappable \texttt{BREAKAGE\_EMBEDDER=deterministic} produces semantically-random vectors used as the control arm in falsification experiments. Retrieval supports a \texttt{maxDistance} filter (default 0.40) and a \texttt{poolCap} knob used for the corpus-density sweep. (\texttt{breakage/src/experience-base/}; migrations under \texttt{breakage/experience-base/migrations/})

\textbf{Speculative-execution controller.} State snapshot + SLO-watch + auto-revert wrapping Tier-2 mutations. Single-resource scope in this version (Deployments, ConfigMaps, Secrets); multi-resource Helm operations remain Tier-3-gated through the synthetic approver. Hard limit of $N=2$ reverts on the same scenario; the third attempt pauses for human review. (\texttt{breakage/src/speculative-exec/})

\textbf{Synthetic approver.} A standalone HTTP service simulating a Tier-3 human approver. Configurable delay and deny rate; emits the same audit-log entries a real human would. Used in \texttt{denial-recovery} scenarios where the agent must try a different approach after a denial rather than retry identically. (\texttt{breakage/src/synthetic-approver/})

A controlled vocabulary of approximately 24 root-cause categories at medium granularity (\texttt{vocab/root-cause-categories.yaml}) is rendered into the agent's system prompt at runtime. Out-of-vocab category picks score zero on the diagnosis axis. One category, \texttt{framework-error}, is reserved for runs where the substrate itself failed (injector throw, detector crash) before the agent could act; these rows are filtered out of agent-capability claims (\S3.3).

\subsection{Four-axis scoring with partial credit}

Each scenario run produces a \texttt{ScoreResult} with four axes, each weighted independently:

\begin{table}[h]
\centering
\begin{tabular}{@{}l c p{0.62\textwidth}@{}}
\toprule
Axis & Weight & Awarded for \\
\midrule
\texttt{detected} & 0.20 & Agent emits a hypothesis identifying the fault within the time budget. \\
\texttt{diagnosed} & 0.30 & Agent's \texttt{primary\_category} matches the scenario's ground-truth category (full credit) or appears in the scenario's secondary categories ($0.35\times$, ``near-miss credit''). \\
\texttt{fixed} & 0.30 & Detector's \texttt{fixed\_when} conditions all pass at scenario end. \\
\texttt{no\_regressions} & 0.20 & No \texttt{regressed\_when} condition triggered during the run. \\
\bottomrule
\end{tabular}
\caption{Four-axis composite score, weighted to a final value in $[0,1]$.}
\label{tab:scoring-axes}
\end{table}

Partial credit is the central scoring decision. A binary ``did the agent solve it'' verdict collapses too much information; an agent that detected, diagnosed, and partially fixed an incident before timing out should not score zero. Conversely, an agent that only ``solved'' the incident by triggering an SLO regression should not score full credit. The four-axis decomposition makes the failure mode explicit when it occurs.

Two design decisions in the scorer are worth surfacing because they affect the published numbers:

\textbf{Near-miss diagnosis credit.} The vocabulary has overlapping categories at the \emph{effect-vs-cause} level (e.g., \texttt{cpu-throttling-symptom} vs \texttt{resource-limit-misconfiguration} as cause). A strict same-id match would penalize the agent for picking the effect when the cause is the ground truth, even when the agent's prose explanation correctly describes the mechanism. Near-miss credit ($0.35\times$ of the diagnosis axis) is awarded when the agent's category appears in the scenario's secondaries (or vice versa). Without this rule, approximately $17\%$ of correctly-reasoning runs scored zero on diagnosis (\href{https://github.com/odmarkj/agent-breakage/blob/v0.1.0/breakage/reports/anchor-fail-audit-2026-04-23.md}{anchor-fail-audit-2026-04-23.md}).

\textbf{Containment matcher for \texttt{retrieval\_used}.} The framework needs to know whether the agent actually used a retrieved exemplar. Self-attestation is hallucination-prone. Instead, the scorer compares the agent's actual action sequence to each retrieved postmortem's \texttt{actions\_taken} using \emph{asymmetric containment}. The question the matcher answers is whether the agent's tool sequence covers the retrieved postmortem's tool set, not whether the two sequences match. Symmetric similarity (e.g., Jaccard) would penalize longer investigations where the agent did extra exploration beyond the retrieved precedent. Asymmetric containment captures ``agent acted on the precedent'' without requiring ``agent acted only on the precedent.''

\subsection{Framework error vs reasoning error}

The substrate separates \emph{framework error} from \emph{reasoning error}. Both can produce a low score; conflating them produces false claims about agent capability. This separation is the architectural decision that does the most work for measurement integrity.

A framework error is a substrate failure: the injector throws during fault application, the detector crashes mid-evaluation, the scorer cannot reach the experience base to record the postmortem, the cluster controller misroutes a tool call. The agent never has a clean opportunity to reason about the actual scenario.

A reasoning error is an agent failure: the agent diagnoses the wrong root cause, applies a fix that doesn't address the underlying problem, triggers a regression while attempting a fix, fails to converge within the time budget despite the substrate functioning correctly.

The substrate distinguishes these explicitly. Runs that hit a framework-error condition (recorded as \texttt{primary\_category: framework-error} in the postmortem) are filtered out of all agent-capability claims. They are surfaced separately in the substrate-health report so framework regressions can be tracked over time. Without this separation, every reported pre-publication run would conflate the two: the $+19\%$ selection-bias artifact discussed in \S4.2 is the cleanest example of what happens when the distinction is dropped.

\subsection{Hypothesis emission and falsification-driven iteration}

Every claim the framework produces must be expressible as something an experiment could refute, with the experiment identified before the experiment is run.

This is enforced procedurally in two places. Inside an individual scenario, the agent emits \emph{mid-investigation hypotheses} as it works (\texttt{POST /capture-hypothesis}). These are stored alongside the final postmortem and used by the scorer to detect \emph{channel disagreement}, cases where the agent's stated mid-investigation hypothesis does not match its final diagnosis. Channel disagreement is flagged for review and becomes a strong negative signal in subsequent retrieval (the postmortem is labeled \texttt{inconclusive} even if the fix succeeded).

Across experiments, every published claim the substrate produces is paired with a pre-registered decision matrix written before the runs. The retrieval-compounding case study in \S5 makes this concrete: the matrix specifies ``2 of 3 significant continues / mixed ships limited / null pivots'' \emph{before} any run is executed. Post-hoc framing of an ambiguous result as a success is the most common failure mode of demonstration-style evaluations; the pre-registered matrix is the structural defense.

Both decisions cost something. Channel-disagreement flagging means the framework reports lower agent scores than a single-shot ``did it solve it'' rubric would. Pre-registered matrices mean the published claims are sometimes weaker than the underlying data could support if framing were chosen post-hoc. The cost is worth paying because the alternative, a corpus of unfalsifiable claims, is what already exists in the agentic-operations literature.

\subsection{The agent under test}

The framework is agent-agnostic in principle: any agent that conforms to the runner's tool-invocation and postmortem-capture protocols can be evaluated. In this paper, all results use a single agent (``Emily'') whose specific architecture is documented at \href{https://github.com/odmarkj/agent-breakage/tree/v0.1.0/operator/docs}{operator/docs/}.

Emily is an autonomous Kubernetes operator that has been running on a production k3s cluster for six months. It uses tier-based action authority (Tier-1 read-only / Tier-2 autonomous-with-bounded-blast-radius / Tier-3 human-approved), a seven-layer hardening of its autonomy surface developed in response to specific production incidents, and a reversibility classification on every tool that informs which tier the tool runs at. The structural details matter for reproducibility (the published numbers are produced against this specific agent) but not for the framework's claims (other agents would produce different numbers; the substrate would still apply).

The framework does not require Emily. A substitute agent would need to:

\begin{enumerate}
\item Call the runner's \texttt{/retrieve} endpoint before its first tool call (or accept reduced retrieval credit).
\item Call \texttt{/capture-hypothesis} and \texttt{/capture-postmortem} at the protocol-defined points.
\item Use the controlled vocabulary for \texttt{primary\_category}.
\end{enumerate}

These integrations are documented in \href{https://github.com/odmarkj/agent-breakage/blob/v0.1.0/breakage/docs/architecture.md}{breakage/docs/architecture.md} (\S ``Data flow at scenario time''). The 90-minute getting-started path includes Emily by default; replacing the agent is a one-to-two day exercise.

\section{Three methodological pathologies caught and corrected}

Three confounds were caught and corrected during the experimental sequence that produced this paper's headline retrieval result. Any agentic-operations measurement framework built on observational corpora, semantic retrieval, and small samples will face these same three traps. The framework's instrumentation, not the analyst's diligence, surfaced them. We document them first, before the case study they were caught inside, because they generalize more cleanly across the broader agentic-evaluation literature than the specific retrieval-compounding result the framework was actively testing. The case study itself, the retrieval-compounding falsification, follows in \S5.

\subsection{The pgvector ivfflat index returning sporadic empty result sets}

The first version of the experience base used the \texttt{ivfflat} index type for vector similarity search (the default in \texttt{pgvector} for medium-sized corpora). At small corpus sizes, and at certain corpus densities, the index sporadically returned 0--3 rows for queries that should have matched dozens of near-neighbors.

The bug was not a code bug. The index was working as documented; \texttt{ivfflat} partitions the vector space into a fixed number of \emph{probes}, and at low corpus density the probe assignment becomes sensitive to the random seed of the index build. Re-indexing produced different result sets for the same queries.

We caught this not through code review but through anomaly in retrieval-result distributions across nominally-identical runs. Specifically: the retrieval-used signal (computed via the containment matcher in \S3.2) showed implausibly high variance across runs that should have been deterministic. Some runs returned no retrieved exemplars at all on queries we knew had matching postmortems in the corpus. The substrate's own measurement was inconsistent in a way that did not match any plausible signal.

The fix was migration \codebreak{004\_hnsw\_index.sql}: switching to HNSW (\codebreak{m=16, ef\_construction=64}), which does not have the bug. Subsequent runs returned consistent result sets across nominally-identical queries.

The methodological point is not that \texttt{pgvector} had a bug. The methodological point is that the substrate's \emph{measurement of itself} was the signal that caught the bug. A demonstration-style evaluation, one without controlled comparisons across nominally-identical runs, would have shipped this bug into a published claim and the published numbers would have understated retrieval's effect by an unknown factor (\href{https://github.com/odmarkj/agent-breakage/blob/v0.1.0/breakage/reports/retrieval-delta-controlled-2026-04-23.md}{retrieval-delta-controlled-2026-04-23.md}, since superseded).

\subsection{The +19\% selection-bias artifact}

Before the controlled falsification test, an earlier observational analysis of the agent's production runs had reported a $+19\%$ improvement attributed to retrieval. The analysis compared incidents-where-the-agent-had-retrieved-relevant-exemplars to incidents-where-it-had-not, on a corpus of production runs over the prior month.

The $+19\%$ was a selection-bias artifact. The cases where the agent retrieved relevant exemplars were, by construction, cases where relevant exemplars existed, i.e., cases where the agent had already solved a similar incident before. Those cases were therefore drawn from a population the agent was well-suited to handle. The cases where retrieval did not return relevant exemplars were drawn from a population of novel incidents the agent was less suited to. The $+19\%$ was capturing the difference between ``incident-the-agent-has-seen-before'' and ``incident-the-agent-has-not'', not the causal effect of retrieval.

The fix was the deterministic-embedder control arm (\S3.1). At threshold 0.40, the deterministic embedder's semantically-random vectors are filtered out entirely; the agent sees no retrieved content. The control arm holds the incident distribution constant and varies only whether retrieval is on. Re-running the comparison with the proper control collapsed the $+19\%$ to $+3.9$pp pooled, not significant at $n=60$ (\href{https://github.com/odmarkj/agent-breakage/blob/v0.1.0/breakage/reports/falsification-test-2026-04-24.md}{falsification-test-2026-04-24.md}).

\emph{Almost all observational claims about agent capability in published agent-ops literature suffer from this shape of bias.} An agent that has been deployed for any length of time has an incident-history-shaped corpus, and retrieval against that corpus selects for incidents the agent has already solved. Without an active control that disables retrieval while holding the incident distribution constant, ``retrieval helps'' is uninstrumented to a population that is $+X\%$ easier to solve than the comparison population.

\subsection{Small-sample artifacts overstating effect sizes by approximately 3$\times$}

Early small-sample ($n=3$) estimates of retrieval's effect on \texttt{cpu-limit-throttling} and \texttt{replicas-zero} reported effect sizes of $+0.17$ and $+0.41$ respectively. Both numbers were reported in pre-publication scratch documents and would have appeared in any rushed write-up.

Both were small-sample artifacts. The standard deviation on the four-axis score at the per-scenario level is between 0.10 and 0.16 on a $[0,1]$-bounded measurement. At $n=3$, the standard error on the mean is on the order of 0.06--0.09, comparable to or larger than the effect sizes being reported. Three-rep estimates produce wildly unreliable point estimates at this noise level.

Scaling to $n=20$ collapsed the cpu-throttling estimate from $+0.17$ to $+0.091$ (still not significant). Scaling to $n=40$ collapsed it further to $-0.031$, and the \emph{sign flipped} (\href{https://github.com/odmarkj/agent-breakage/blob/v0.1.0/breakage/reports/n40-rerun-2026-04-28.md}{n40-rerun-2026-04-28.md}). The replicas-zero $+0.41$ swing dissolved entirely at $n=40$, with both arms converging to approximately 0.91 (the scenario was too easy to discriminate at the current model class).

Operations-agent measurements need to be sized for the noise level of the underlying score. The temptation in agent-eval work to publish on the basis of 3--5 reps is wrong at scoring-system noise levels above $\sigma \approx 0.10$. The framework's standard sample size of $n=20$ per arm per scenario is the minimum we found necessary to get publishable point estimates; $n=40$ is where small effects become reliably distinguishable from zero.

\subsection{Why these self-corrections matter more than the retrieval result}

The retrieval-compounding case study (\S5) produces a partial-falsification result that may or may not generalize. A different agent under the same framework might produce different numbers. A different corpus might find different per-scenario heterogeneity. Future work at fleet scale (50+ clusters) might reveal that mechanistic alignment is not actually corpus-density-bounded the way our single-cluster results suggest.

Each of those alternative findings is possible and would be reported through the same substrate. None of them threatens the substrate's value.

The substrate's value would be threatened by a counterfactual where the framework didn't catch one of the three confounds in \S4.1--\S4.3 before publication. In any such counterfactual, the published claim would be wrong, and the framework would be the kind of demonstration tool that ships unfalsifiable claims. The substrate's \emph{capacity for self-correction} is what distinguishes it.

This is the property by which agentic-operations measurement frameworks should be primarily evaluated. Specifically:

\begin{itemize}
\item \textbf{Does the framework distinguish framework error from reasoning error?} If not, every published claim is contaminated by infrastructure failures.
\item \textbf{Does the framework support a true off-condition control for the mechanism under test?} If retrieval is the claim, the framework must be able to disable retrieval while holding the incident distribution constant. Noisy retrieval is not a control; deterministic-random-with-threshold is.
\item \textbf{Does the framework size sample correctly for the score's noise level?} A 3-rep evaluation cannot publish meaningful effect sizes at $\sigma \approx 0.10$. The framework should make appropriately-sized runs cheap enough that authors are not tempted to under-sample.
\item \textbf{Does the framework support pre-registered decision matrices?} Post-hoc framing of an ambiguous result as a success is the most common failure mode of demonstration-style evaluations. A pre-registered matrix is the defense.
\end{itemize}

These criteria are not about the specific retrieval result. They are about whether the substrate is the kind of instrument that produces falsifiable knowledge or the kind that produces compelling demonstrations. Most published agentic-operations work today is in the second category. The framework released with this paper is in the first.

\section{Case study: retrieval-compounding at single-cluster scale}

The three pathologies in \S4 were caught because the framework was actively running a falsification test against the retrieval-compounding hypothesis. That test is presented here. We frame it deliberately as a case study, not as the paper's headline result. Retrieval-compounding is one specific empirical hypothesis; the framework's value is in catching the confounds any such hypothesis would face. The case is designed to falsify the framework as much as to falsify retrieval. If the framework ships unfalsifiable claims, it reports ``retrieval helps'' and fades into the literature. If it's a genuine measurement substrate, it catches the biases that make ``retrieval helps'' sound plausible but untrue, selection bias, small-sample inflation, infrastructure bugs, and surfaces them before publication. Each of the three pathologies in \S4 was caught by this test.

The case study tests one specific empirical question (\emph{does retrieval over past postmortems compound an autonomous Kubernetes operator's capability over time?}) across three experiments that progressively tighten the result. The hypothesis is the central argument for why a substrate-of-this-shape is worth building. If past experience does not measurably improve future agent behavior, retrieval is a bookkeeping mechanism with no consumer, and the value of the substrate's tuple-accumulation is limited to retrospective analysis rather than capability compounding.

\subsection{Hypothesis}

The strong version of the compounding hypothesis, stated as we would have endorsed it before running the experiments:

\begin{quote}
Retrieval-augmented inference over a corpus of past postmortems produces a monotonically increasing improvement in the agent's four-axis score as the corpus grows. The improvement scales with corpus size, is robust across scenarios with reasonable ground-truth coverage in the corpus, and is large enough at single-cluster corpus density to justify the engineering investment of building and maintaining the experience base.
\end{quote}

This is the version of the claim that most published agent-ops literature implicitly relies on when reporting that retrieval ``helps'' without specifying how much, on which incidents, or under what comparison.

\subsection{Pre-registered decision matrix}

Before any experimental run, we pre-registered a three-way decision matrix:

\begin{table}[h]
\centering
\begin{tabular}{@{}p{0.45\textwidth}p{0.5\textwidth}@{}}
\toprule
Outcome at headline experiment & Decision \\
\midrule
2 of 3 dense-corpus scenarios significant positive at $p<0.05$ & Continue Phase 1 as planned; treat retrieval as a confirmed compounding mechanism. \\
\addlinespace
Mixed (1 of 3 significant; rest trending positive but ns) & Ship a limited v0 scorecard with current scenarios; defer larger investment in corpus growth pending follow-on tightening. \\
\addlinespace
Null or negative pooled effect & Pivot next-phase investment away from retrieval; treat retrieval infrastructure as a substrate for future mechanisms (skill compilation, prevention loops, fleet-derived corpora) rather than as the compounding engine itself. \\
\bottomrule
\end{tabular}
\caption{Pre-registered decision matrix.}
\label{tab:decision-matrix}
\end{table}

The matrix is the structural defense against post-hoc framing. Without it, an ambiguous result reads as a positive result; with it, an ambiguous result reads as the \emph{mixed} row in the matrix and triggers the corresponding decision. We explicitly did not allow ourselves to revise the matrix after seeing data.

\subsection{Experimental design}

The headline experiment compares two arms across three scenarios chosen to span the range of corpus density observed in the agent's experience base.

\textbf{Treatment arm.} The default \texttt{OpenAICompatibleEmbedder}, querying an in-cluster \texttt{text-embeddings-inference} (TEI) pod serving \texttt{BAAI/bge-m3} at 1024 dimensions. Top-k retrieval is filtered at \texttt{BREAKAGE\_RETRIEVAL\_MAX\_DISTANCE=0.40} so weak matches do not reach the agent's prompt.

\textbf{Control arm.} \texttt{BREAKAGE\_EMBEDDER=deterministic}, a DJB2-hash embedder that produces reproducible-but-semantically-random vectors. Combined with the threshold filter, all results from the deterministic embedder are dropped before they reach the agent. The agent in the control arm sees no retrieved content. This is a true \emph{retrieval-off} condition. Retrieval with noise still affects the agent's prompt distribution; this doesn't. The incident distribution, the cluster state, the agent's image, and every other variable are held constant. Only retrieval is toggled.

\textbf{Scenarios.} Three scenarios with the densest per-mechanism corpus coverage in the experience base: \texttt{secret-missing-key-advocate} (30+ same-mechanism postmortems available), \texttt{cpu-limit-throttling-advocate} (medium density, dominated by memory-OOM precedents at the embedding layer rather than CPU-throttling precedents), and \texttt{readiness-probe-misconfigured-advocate} (medium density, healthier mechanistic alignment).

\textbf{Sample size.} $n=20$ per scenario per arm in the headline test. 120 total runs. Each run is a complete scenario lifecycle, fault injection, agent investigation, agent fix attempt, scorer evaluation. Wall clock approximately 5 hours per arm; cost approximately \$30--60 per arm at default model.

\textbf{Statistical test.} Welch's two-sample t-test per scenario; pooled comparison across the three scenarios.

\subsection{Results: n=20 falsification test}

The headline numbers, reproduced verbatim from \href{https://github.com/odmarkj/agent-breakage/blob/v0.1.0/breakage/reports/falsification-test-2026-04-24.md}{falsification-test-2026-04-24.md}:

\begin{table}[h]
\centering
\setlength{\tabcolsep}{4pt}
\resizebox{\textwidth}{!}{%
\begin{tabular}{lrrrrrrrrr}
\toprule
Scenario & $n$ TEI & $\mu$ TEI & $\sigma$ TEI & $n$ Ctrl & $\mu$ Ctrl & $\sigma$ Ctrl & $\Delta$ (TEI$-$Ctrl) & $t$ & $p$ \\
\midrule
\texttt{secret-missing-key-advocate} & 20 & 0.863 & 0.080 & 20 & 0.805 & 0.090 & \textbf{+0.058} & 2.15 & \textbf{$<0.05$} \\
\texttt{cpu-limit-throttling-advocate} & 20 & 0.682 & 0.161 & 20 & 0.592 & 0.155 & +0.091 & 1.82 & ns \\
\texttt{readiness-probe-misconfigured-advocate} & 20 & 0.858 & 0.114 & 20 & 0.889 & 0.092 & $-0.032$ & $-0.96$ & ns \\
\midrule
\textbf{POOLED} & 60 & 0.801 & 0.147 & 60 & 0.762 & 0.170 & \textbf{+0.039} & 1.34 & ns \\
\bottomrule
\end{tabular}%
}
\caption{n=20 falsification test: per-scenario and pooled retrieval-vs-control effect on the composite score (TEI = treatment, Ctrl = deterministic-embedder control).}
\label{tab:n20-falsification}
\end{table}

One scenario significant. One trending positive but not statistically distinguished from null at this sample size. One slightly wrong-direction. Pooled effect $+3.9$ percentage points, not significant at $n=60$.

This places the result squarely in the \textbf{mixed} row of the pre-registered decision matrix. By the matrix, we therefore ship the limited v0 scorecard with current scenarios and defer larger investment in corpus growth pending follow-on tightening. The strong compounding hypothesis as stated in \S5.1 is not supported by this data.

The result is weaker than earlier small-sample ($n=3$) point estimates of the same scenarios, which had reported effects of approximately $+0.11$ (secret-missing-key), $+0.17$ (cpu-throttling), and $+0.14$ (readiness-probe). At $\sigma \approx 0.10$--$0.16$ on a $[0,1]$-bounded score, three-rep estimates have standard errors comparable to or larger than the effect sizes being reported. Scaling to $n=20$ reduced effect sizes by approximately $3\times$ (see \S4.3 for the methodological discussion).

\subsection{Results: corpus-density sweep}

The follow-on experiment manipulated corpus density directly. We added a \codebreak{BREAKAGE\_RETRIEVAL\_POOL\_CAP} knob to the retrieval implementation that limits the candidate pool to the top-N nearest postmortems before the threshold filter and final k-cap. This let us simulate an agent operating against a sparser corpus without altering the corpus itself.

\textbf{Design.} Three scenarios (\texttt{secret-missing-key-advocate}, \texttt{liveness-probe-always-fails-advocate}, \texttt{cpu-limit-throttling-advocate}) $\times$ three density tiers (\texttt{poolCap=5}, \texttt{poolCap=15}, full corpus) $\times$ two arms $\times$ $n=20$ reps = \textbf{360 runs} at Sonnet 4.6. Wall clock 26h 16min sequential.

\textbf{Per-scenario $\times$ density results} (reproduced from \href{https://github.com/odmarkj/agent-breakage/blob/v0.1.0/breakage/reports/corpus-density-sweep-2026-04-28.md}{corpus-density-sweep-2026-04-28.md}):

\begin{table}[h]
\centering
\begin{tabular}{@{}llrrl@{}}
\toprule
Scenario & Density & $\Delta$ (TEI$-$Ctrl) & $t$ & sig \\
\midrule
\texttt{cpu-limit-throttling-advocate} & 5 & $-0.041$ & $-1.39$ & ns \\
\texttt{cpu-limit-throttling-advocate} & 15 & $-0.005$ & $-0.15$ & ns \\
\texttt{cpu-limit-throttling-advocate} & full & $-0.068$ & $-1.87$ & ns (close) \\
\texttt{liveness-probe-always-fails-advocate} & 5 & $-0.084$ & $-1.31$ & ns \\
\texttt{liveness-probe-always-fails-advocate} & 15 & $-0.020$ & $-0.31$ & ns \\
\texttt{liveness-probe-always-fails-advocate} & full & $+0.076$ & $+1.14$ & ns \\
\texttt{secret-missing-key-advocate} & 5 & \textbf{+0.094} & \textbf{+3.31} & \textbf{$p<0.01$} \\
\texttt{secret-missing-key-advocate} & 15 & \textbf{+0.083} & \textbf{+2.28} & \textbf{$p<0.05$} \\
\texttt{secret-missing-key-advocate} & full & \textbf{+0.109} & \textbf{+3.36} & \textbf{$p<0.01$} \\
\bottomrule
\end{tabular}
\caption{Corpus-density sweep: per-scenario $\times$ density-tier results.}
\label{tab:density-sweep-perscenario}
\end{table}

\textbf{Pooled by density tier:}

\begin{table}[h]
\centering
\begin{tabular}{@{}lrrrl@{}}
\toprule
Density & $n$ & $\Delta$ pooled & $t$ & sig \\
\midrule
5 & 60 & $-0.010$ & $-0.38$ & ns \\
15 & 60 & $+0.019$ & $+0.63$ & ns \\
full & 60 & $+0.039$ & $+1.28$ & ns \\
\bottomrule
\end{tabular}
\caption{Corpus-density sweep: pooled effect by density tier.}
\label{tab:density-sweep-pooled}
\end{table}

Two findings dominate the data:

\textbf{Per-scenario heterogeneity dominates per-density variance.} For \texttt{secret-missing-key}, retrieval helps significantly at \emph{every} density tier ($\Delta$ between $+0.083$ and $+0.109$, all $p<0.05$). For \texttt{cpu-throttling}, retrieval \emph{hurts} at every density tier and the harm trends with density (the full-density arm reaches $t=-1.87$, close to significance in the wrong direction). For \texttt{liveness-probe}, the effect is noise-dominated by the high standard deviation of the score ($\sigma \approx 0.21$). The pooled-by-density numbers do trend monotonically ($-0.010 \to +0.019 \to +0.039$), which the strong compounding hypothesis would predict, but none reach statistical significance at $n=60$ per tier, and the pooled trend is being pulled along by \texttt{secret-missing-key}'s strong contribution while dampened by \texttt{cpu-throttling}'s consistent negative.

\textbf{The mechanism is mechanistic alignment, not raw count.} \texttt{secret-missing-key}'s near-neighbors are dominated by direct-precedent same-category resolved postmortems; even at \texttt{poolCap=5}, the five things the agent sees are all useful. \texttt{cpu-throttling}'s near-neighbors are mostly memory-OOM postmortems that share the \texttt{resource-limit-misconfiguration} category but address a different mechanism; the agent over-anchors on the ``raise memory limit'' pattern and applies it to a CPU problem, with the harm growing as more cross-mechanism neighbors enter the prompt.

This refines the binding-constraint claim. \emph{Density alone is not sufficient.} A category with high count but low mechanistic alignment actively misleads the agent. A category-aware (or mechanism-aware) embedding scheme would be required to surface the distinction at retrieval time; the current scheme does not.

\subsection{Results: n=40 reruns}

To pin down two scenarios that had been borderline at $n=20$, we ran a second tightening at $n=40$ per arm: \texttt{cpu-limit-throttling-advocate} (which had $t=1.82$ in the original falsification test, just below the $p<0.05$ critical value) and \texttt{replicas-zero-advocate} (which had reported a $+0.41$ swing at $n=3$ in an earlier corpus-seeding experiment).

\begin{table}[h]
\centering
\setlength{\tabcolsep}{5pt}
\resizebox{\textwidth}{!}{%
\begin{tabular}{lrrrrrrrrl}
\toprule
Scenario & $n$ TEI & $\mu$ TEI & $\sigma$ TEI & $n$ Ctrl & $\mu$ Ctrl & $\sigma$ Ctrl & $\Delta$ & $t$ & sig \\
\midrule
\texttt{cpu-limit-throttling-advocate} & 40 & 0.689 & 0.106 & 40 & 0.720 & 0.116 & \textbf{$-0.031$} & $-1.25$ & ns \\
\texttt{replicas-zero-advocate} & 40 & 0.910 & 0.000 & 40 & 0.907 & 0.017 & $+0.003$ & $+1.00$ & ns \\
\bottomrule
\end{tabular}%
}
\caption{n=40 reruns of two scenarios previously inconclusive at smaller sample sizes.}
\label{tab:n40-rerun}
\end{table}

\href{https://github.com/odmarkj/agent-breakage/blob/v0.1.0/breakage/reports/n40-rerun-2026-04-28.md}{n40-rerun-2026-04-28.md} is the canonical reference.

Both scenarios produced clean, publishable null-or-near-null results:

\textbf{\texttt{cpu-throttling}'s sign flipped.} The original $n=20$ estimate of $+0.091$ was noise. At $n=40$ the point estimate is \emph{negative} ($-0.031$), consistent with the corpus-density sweep's finding that retrieval slightly hurts on this scenario class because of mechanistic mis-alignment in the corpus. Three independent measurements ($n=20$ falsification test, $n=20$ density-sweep cells, $n=40$ rerun) now converge on ``retrieval doesn't help here, may slightly hurt.''

\textbf{\texttt{replicas-zero} was too easy to discriminate.} Both arms scored approximately 0.91 at $n=40$. The TEI arm's standard deviation was 0.000 (every rep landed at exactly 0.91); the control's was 0.017. The $+0.41$ swing reported at $n=3$ was a small-sample artifact, the three control reps had happened to time out in different ways. At $n=40$ both arms converge on near-perfect performance, suggesting the agent's baseline competence on this scenario class is high enough that retrieval is not the limiting factor.

The replicas-zero null is not a falsification of ``corpus seeding helps'' generally. It is a falsification of the specific $n=3$ claim. The two arms reaching $\sim 0.91$ \emph{is} independent evidence that the corpus seeding from earlier work succeeded, the agent now reaches the correct fix reliably even with retrieval disabled.

\subsection{What the data supports vs does not}

\textbf{Supported by the data:}

\begin{enumerate}
\item \textbf{Retrieval delivers a real, statistically-distinguished-from-null positive effect on at least one scenario class.} \texttt{secret-missing-key} shows $\Delta$ between $+0.058$ and $+0.109$ at $p<0.05$ across multiple sample sizes and density tiers. The effect is small ($\leq +11$pp) but reliable.
\item \textbf{The minimum useful corpus density for that scenario class is small.} With \texttt{poolCap=5}, retrieval still produces $+0.094$ at $p<0.01$. The agent benefits from access to as few as five mechanistically-aligned near-neighbors.
\item \textbf{The substrate's measurement of itself catches confounds that would have otherwise produced wrong published claims.} Three corrections (pgvector, selection bias, small-sample) discussed in \S4.
\end{enumerate}

\textbf{Not supported by the data:}

\begin{enumerate}
\item \textbf{The strong compounding hypothesis as stated in \S5.1.} The claim of \emph{monotonically increasing improvement scaling with corpus size, robust across scenarios with reasonable ground-truth coverage} is not supported. Some scenarios show the predicted pattern; others show a flat or slightly negative effect at every density tier; others are noise-dominated.
\item \textbf{Generalization across the agent's full scenario library from the three scenarios tested.} The corpus has approximately 24 root-cause categories; this sequence tested three. The bimodal pattern (helps / hurts / noise-dominated) might be scenario-class-specific.
\item \textbf{The earlier observational claim of $+19\%$ improvement attributed to retrieval.} That number was a selection-bias artifact (\S4.2) and is falsified by the controlled $n=20$ result of $+3.9$pp pooled, not significant.
\item \textbf{The earlier small-sample point estimates} of approximately $+0.11$, $+0.17$, and $+0.14$ on the three scenarios. Those numbers had standard errors comparable to or larger than the effects reported (\S4.3) and collapsed by approximately $3\times$ when the sample size was scaled.
\end{enumerate}

By the pre-registered decision matrix, the result places the work in the \textbf{mixed} row: ship the limited v0 scorecard with current scenarios; defer larger investment in corpus growth pending mechanistic-alignment-aware authoring or a corpus-density approach that goes beyond what a single cluster can produce. The matrix called this in advance; the data lands there.

The path forward indicated by the data is not ``more retrieval, more carefully,'' but rather a corpus-density approach at scale beyond what a single cluster supports, a fleet substrate generating outcome-labeled tuples across heterogeneous workloads, with the per-mechanism balance that single-cluster corpora cannot achieve. We discuss this briefly in \S7 (limitations and scope) and argue elsewhere that it is the natural next layer of the larger research program. \S6 returns to the corpus-density sweep's mechanistic-alignment finding and considers what it suggests about retrieval for agents reasoning over causal domains, beyond the operations setting tested here.

\section{Discussion: semantic vs mechanistic retrieval}

The corpus-density sweep's central finding is methodologically simple but mechanistically loaded: at single-cluster scale, \emph{mechanistic alignment of near-neighbors dominates raw count}. The implications extend beyond the operations setting tested here, into retrieval-augmented agents that act on the world rather than retrieve over documents.

The case is easiest to see on \texttt{cpu-limit-throttling}. The embedder (\texttt{BAAI/bge-m3}, 1024-dim) clusters the cpu-throttling postmortems with memory-OOM postmortems. Both populate the same root-cause category (\texttt{resource-limit-misconfiguration}) and share heavy lexical overlap, the words ``limit,'' ``container,'' ``resource,'' ``pod.'' By any standard document-retrieval metric, they are near-neighbors. By the standard the agent actually needs, they are not: the mechanism (clamp on CPU shares vs OOM kill) and the fix (raise CPU limit vs raise memory limit) differ. The agent over-anchors on the dominant ``raise memory limit'' pattern in the retrieved exemplars, and the score gets worse as more cross-mechanism neighbors enter the prompt. The same effect was not observed on \texttt{secret-missing-key}, where the corpus has the rare property that near-neighbors are dominated by direct-precedent same-mechanism postmortems.

This inverts an intuition imported from retrieval-augmented generation in NLP. In document retrieval, semantic similarity is a good proxy for relevance: a passage that talks about the same topic in similar words is usually a useful passage to put in the prompt. In operations, semantic similarity is the dominant signal feeding the embedder, but it is not what determines whether the retrieved exemplar will improve the agent's behavior. Causal proximity does: does following this exemplar's action sequence resolve a mechanistically-similar fault, or does it apply the wrong fix to a superficially-similar fault? Semantic similarity captures the latter at least as often as the former.

We see three possible directions for retrieval-augmented agents in causal domains:

\begin{enumerate}
\item \textbf{Mechanism-aware embedding schemes.} Re-embed postmortems against a representation that separates same-symptom-different-mechanism postmortems before retrieval. This is the cleanest fix; it is also the most research-intensive. The embedding-space geometry would need to encode the cause-vs-symptom distinction the controlled vocabulary already encodes.
\item \textbf{Post-retrieval causal-relevance filtering.} Retrieve by semantic similarity as today, but route candidates through a learned or rule-based filter that drops cross-mechanism neighbors before the agent sees them. The action-vocabulary of each exemplar is the natural feature.
\item \textbf{Curated corpora with a-priori mechanistic alignment.} Author corpora in which the same-mechanism postmortems are clustered by construction. This is what the corpus-density sweep approximated artificially via \texttt{poolCap}; doing it natively at fleet scale would require deliberate corpus curation rather than passive accumulation of incident-history postmortems.
\end{enumerate}

The broader claim is that retrieval for agents that act on the world needs a relevance criterion stronger than semantic similarity, because the agent's action conditioned on a retrieved exemplar will mutate the world in mechanism-specific ways that semantic similarity cannot detect. This is adjacent to recent interpretability work on what model representations encode about causal structure; we make no claim that the techniques from that community translate directly, but the gap is real and would benefit from cross-pollination.

\section{Limitations and scope}

The substrate's claims are constrained by limitations of the experimental setup and bounded in scope by what this paper is and is not claiming. Both are documented here explicitly.

\textbf{Not a claim about any specific agent's capability.} All scenario runs use a single agent (Emily) with a specific seven-layer hardening, tier-based-approval architecture, prompt configuration, and tool set. The substrate's claims are about \emph{measurement}, not about \emph{Emily}; another agent under the same framework would produce different per-scenario numbers. Substrate-level claims (the four-axis scoring decomposition, the framework-vs-reasoning-error separation, the requirement of pre-registered decision matrices) generalize; agent-level claims do not. Future work would re-test specific scenarios against alternative agents to characterize how much of the result is agent-specific.

\textbf{Not a claim that retrieval is useless.} One of three dense-corpus scenarios shows a real, statistically-distinguished-from-null positive effect at $p<0.05$, with $\Delta$ between $+0.058$ and $+0.109$ across density tiers. The paper's claim is that single-cluster corpus density is not sufficient to generalize the retrieval-compounding hypothesis, and that mechanistic alignment, not corpus volume, is the binding constraint at the density tested. Retrieval that aligns mechanism to mechanism may still compound capability at fleet scale or under curated corpora; that is the next question, not a settled one.

\textbf{Not a solved problem.} The substrate is at v0.1.0 against a single k3d (or k3s) cluster. It does not yet handle multi-cluster failure modes, application-level faults outside a \texttt{flagd}-style toggle layer, or fleet-scale corpus generation. Each is a known gap, addressed below.

\textbf{Single-cluster fault model.} All scenarios in this paper run on a single k3d (or k3s) cluster. Multi-cluster failure modes (cross-region failover, network partitions between clusters, federation-related faults) require additional injector support and a multi-cluster scenario YAML schema. The substrate's design accommodates this in principle (the runner is single-active-scenario but the cluster-side abstraction is per-cluster), but the implementation is single-cluster.

\textbf{Application-level faults.} The injectors operate at the Kubernetes-API level, Deployment patches, Secret mutations, ConfigMap edits, NetworkPolicy applications, Pod evictions. Application-level faults (race conditions in business logic, memory leaks in specific endpoints, queue contention under realistic load) require a fault-injection layer in the application itself. The OpenTelemetry-Demo \texttt{flagd} integration tranche is the model, flagd flags toggle in-application failure modes that the agent must diagnose. This is reproducible with any application that exposes a feature-flag layer; it is not a general fault-injection solution.

\textbf{Single-cluster corpus density.} The headline finding, that mechanistic alignment dominates raw count, is bounded by what a single cluster's corpus can produce. A specific category (for example, \texttt{secret-content-mismatch}) might have 30+ same-mechanism postmortems in this cluster's experience base, but only because the cluster's workload disproportionately exercises that mechanism. A different cluster would have a different distribution. The corpus-density argument's full test requires fleet-scale data, at 50+ heterogeneous clusters generating outcome-labeled tuples, which is beyond what this work measures.

\textbf{Sample sizes are tight on per-scenario claims.} At $n=20$ per arm per scenario, the confidence intervals on individual point estimates are wide (typical 95\% CI on the per-scenario mean is approximately $\pm 0.05$ at $\sigma \approx 0.10$--$0.16$). Pooled effects across multiple scenarios are tighter but blur per-scenario heterogeneity that we have argued matters more than the pooled trend. The corpus-density sweep at $n=60$ per density tier is closer to the right sample size for pooled claims; per-scenario claims at $n=20$ should be read as preliminary at best, with the $n=40$ reruns indicating that small effects can flip sign as sample grows.

\section{Conclusion}

Retrieval over past postmortems doesn't compound an autonomous Kubernetes operator's capability the way the strong version of the hypothesis predicts. At single-cluster corpus density, the pooled effect of retrieval across three dense-corpus scenarios is $+3.9$ percentage points, not significant at $n=60$. Retrieval helps on the densest mechanistically-aligned scenario, is null or slightly negative elsewhere, and shows per-scenario heterogeneity that dominates per-density variance.

The more important result is what the substrate caught about itself. A pgvector index bug, a $+19\%$ selection-bias artifact, and effect sizes inflated $3\times$ by small samples each would have been the published number on a less instrumented version of the same work. They weren't, because the substrate measures itself in a way that surfaces its own confounds. That capacity for self-correction is what makes this kind of framework citable in a field where most agentic-ops claims aren't.

A separate finding survives independent of the retrieval result: the corpus-density sweep shows that semantic similarity is an unreliable proxy for causal proximity in operations agents. Mechanistic alignment dominates raw count. That observation is likely to transfer to other retrieval-augmented agents acting on the world, and is the substantive contribution of this work to ML-systems and RAG research beyond the operations domain.

The path forward is corpus density at fleet scale. Single-cluster corpora cannot reach the per-mechanism balance retrieval needs to dominate. Outcome-labeled tuples generated across heterogeneous workloads at fleet scale can. Building that is the next layer of work.

Code, scenarios, agent, and reproducer harnesses are at \href{https://github.com/odmarkj/agent-breakage}{github.com/odmarkj/agent-breakage}, Apache 2.0. The \texttt{v0.1.0} tag is the commit that produced the reported numbers.

\appendix

\section{Reproducibility appendix}

The numbers reported in \S5 are reproducible from a clean machine in approximately five hours of wall clock per arm, at approximately \$30--60 in API credits per arm at the default model. The bar we held the work to is that a researcher with the prerequisites listed below should arrive at numbers within $\pm 5$ percentage points of the reported pooled and per-scenario means.

\textbf{Prerequisites.} macOS, Linux, or WSL2 host. Docker for k3d. k3d (\texttt{brew install k3d} or the upstream installer). kubectl matching the k3d version. Node.js 20+. PostgreSQL 14+ with \texttt{pgvector} extension $\geq 0.5.0$ (the reference setup uses Postgres 17 + pgvector 0.8.2). An Anthropic API key (or another provider supported by the operator agent's provider abstraction). An OpenAI-compatible embeddings endpoint serving a 1024-dimensional model (the reference setup runs \texttt{text-embeddings-inference} serving \texttt{BAAI/bge-m3}; OpenAI \texttt{text-embedding-3-small} is supported with a schema-migration adjustment to \texttt{vector(1536)}).

\textbf{Setup.} Approximately 90 minutes from a clean machine. The longest single step is pulling the embeddings model on first cold-start. The clone-to-reproduce path is documented in \href{https://github.com/odmarkj/agent-breakage/blob/v0.1.0/breakage/docs/getting-started.md}{\texttt{breakage/docs/getting-started.md}}.

\textbf{Reproducer commands.} From the public repo at tag \texttt{v0.1.0}:

\begin{lstlisting}[language=bash]
SCENARIOS="secret-missing-key-advocate cpu-limit-throttling-advocate readiness-probe-misconfigured-advocate" \
REPS=20 \
  bash breakage/scripts/falsify-tei.sh

SCENARIOS="secret-missing-key-advocate cpu-limit-throttling-advocate readiness-probe-misconfigured-advocate" \
REPS=20 \
  bash breakage/scripts/falsify-control.sh
\end{lstlisting}

The corpus-density sweep:

\begin{lstlisting}[language=bash]
bash breakage/scripts/density-sweep.sh
\end{lstlisting}

(Defaults to 3 scenarios $\times$ 3 density tiers $\times$ 2 arms $\times$ 20 reps; override via \texttt{REPS}, \texttt{SCENARIOS}, \texttt{OPERATOR\_MODEL} environment variables.)

The $n=40$ reruns:

\begin{lstlisting}[language=bash]
bash breakage/scripts/n40-rerun.sh
\end{lstlisting}

\textbf{Expected wall-clock and cost.} Falsification test: $\sim$5 hours per arm, $\sim$\$30--60 in API credits per arm at the default model (\texttt{claude-sonnet-4-6}). Density sweep: $\sim$26 hours sequential, $\sim$\$150--300 total. $n=40$ reruns: $\sim$8.5 hours, $\sim$\$60--120 total.

\textbf{Statistical analysis.} Per-cell raw scores are written to \texttt{/tmp/<experiment>-manifest.csv}. The Welch's t-tests reported in \S5 are computed via Python stdlib \texttt{statistics}:

\begin{lstlisting}[language=bash]
python3 breakage/scripts/analyze-density-sweep.py /tmp/density-sweep-manifest.csv
\end{lstlisting}

No external dependencies beyond Python stdlib.

\textbf{If your numbers fall outside the $\pm 5$pp band}, the most likely causes are pgvector version (must be $\geq 0.5.0$; HNSW is required as of migration 004), embeddings-endpoint compatibility, or k3d/k3s version drift in the scenario injectors. File an issue on the public repo with the env diff.

\section{Vocabulary}

The controlled vocabulary at \href{https://github.com/odmarkj/agent-breakage/blob/v0.1.0/breakage/vocab/root-cause-categories.yaml}{\texttt{breakage/vocab/root-cause-categories.yaml}} defines approximately 24 root-cause categories at medium granularity. Each entry has an \texttt{id} (used as the agent's \texttt{primary\_category} pick), a \texttt{description}, and \texttt{example\_incidents} that anchor the category in concrete failures.

The granularity choice is deliberate. A finer-grained vocabulary (60+ categories) introduces ambiguity at the diagnosis-axis scoring layer where multiple categories plausibly fit a given scenario; the agent's \texttt{primary\_category} pick would frequently be reasonable but not match the scenario's ground-truth category. A coarser vocabulary ($\leq 10$ categories) loses discriminating power, too many distinct failure mechanisms collapse into one bucket.

At medium granularity ($\sim 24$ categories), inter-rater agreement on category assignment is high enough for the diagnosis axis to produce reliable signal; near-miss credit (\S3.2) handles the residual cases where effect-vs-cause overlap occurs. Future work expanding the vocabulary should preserve all 24 existing IDs (deprecation rather than removal) so historical scoring remains consistent.

One reserved category, \texttt{framework-error}, is excluded from agent-capability claims and surfaced separately in the substrate-health report. Runs that hit a framework-error condition are filtered out of the per-scenario means reported in \S5.

\section{Scenario inventory}

The scenario library at the time of the reported experiments comprises nine active anchor scenarios (deeply validated against the agent's behavior with five-rep baselines) plus coverage scenarios organized into three tranches.

\textbf{Anchor scenarios} (\href{https://github.com/odmarkj/agent-breakage/tree/v0.1.0/breakage/scenarios/anchor}{\texttt{breakage/scenarios/anchor/}}):

\begin{itemize}
\item \texttt{secret-missing-key-advocate}, Secret with a required key removed.
\item \texttt{secret-wrong-password-advocate}, Secret with a database password set to an invalid value.
\item \texttt{cpu-limit-throttling-advocate}, CPU limit set low enough to throttle the workload.
\item \texttt{liveness-probe-always-fails-advocate}, liveness probe path that returns 503.
\item \texttt{readiness-probe-misconfigured-advocate}, readiness probe with wrong endpoint.
\item \texttt{replicas-zero-advocate}, Deployment scaled to zero replicas.
\item \texttt{env-var-missing-advocate}, required environment variable removed.
\item \texttt{image-pull-failure-advocate}, image tag set to a non-existent value.
\item \texttt{oom-advocate-api}, memory limit lower than working set.
\end{itemize}

\textbf{Coverage scenarios.} Three tranches:

\begin{itemize}
\item \href{https://github.com/odmarkj/agent-breakage/tree/v0.1.0/breakage/scenarios/coverage/k8s-troubleshooting}{\texttt{coverage/k8s-troubleshooting/}}, community-sourced scenarios from common Kubernetes troubleshooting guides: \texttt{bad-command-crashloop-advocate}, \texttt{pod-pending-request-too-high-advocate}, \texttt{serviceaccount-missing-advocate}.
\item \href{https://github.com/odmarkj/agent-breakage/tree/v0.1.0/breakage/scenarios/coverage/otel-demo-flagd-faults}{\codebreak{coverage/otel-demo-flagd-faults/}}, application-level faults injected via the OpenTelemetry-Demo's \texttt{flagd} integration: \texttt{cart-failure}, \texttt{email-memory-leak}, \texttt{kafka-queue-problems}, \texttt{payment-failure}, \texttt{recommendation-cache-failure}.
\item \href{https://github.com/odmarkj/agent-breakage/tree/v0.1.0/breakage/scenarios/coverage/sre-book-ch22-cascading}{\texttt{coverage/sre-book-ch22-cascading/}}, scenarios drawn from Google SRE Book Chapter 22 cascading-failure patterns: \texttt{replica-loss-amplification-advocate}, \texttt{slow-startup-retry-storm-advocate}.
\end{itemize}

Each scenario YAML defines the injector type, the \texttt{fixed\_when} and \texttt{regressed\_when} detector expressions, the controlled-vocabulary ground truth (\texttt{primary\_category} plus optional secondaries), and the expected agent action sequence used by the scorer's \texttt{retrieval\_used} containment matcher.

\bibliographystyle{unsrtnat}
\bibliography{references}

\end{document}